\documentclass[11pt]{article}
\usepackage{hyperref}
\pdfoutput=1
\begin{document}
\title{Dynamically forced cells of a viscoelastic fluid over an array of rollers}
\author{Bin Liu$^1$, Jun Zhang$^{1,2}$, and Michael Shelley$^{1}$\\
\\\vspace{6pt} $^1$ Applied Mathematics Laboratory, Courant Institute of Mathematical Sciences, 
\\New York University, New York, NY 10012, USA
\\ $^2$Department of Physics, New York University, New York, NY 10003, USA}
\maketitle
%% The abstract (in this file, and that submitted as text to arXiv) should include the exact phrase
%% "fluid dynamics video" or "fluid dynamics videos"
\begin{abstract}
Our fluid dynamics video shows the response of a layer of viscoelastic fluid to an array of four-roll mills steadily rotating underneath. When the relaxation time of the fluid is sufficiently long, the fluid divides into ``cells" with a convex free surface above the site of each roller.  This is reminiscent of the rod-climbing effect. On this relaxation time-scale, the flow also transitions from being initially Newtonian-like to one where the fluids' elasticity plays a dynamical role: The fluid cells oscillate with regularity in position and shape on a timescale much longer than the relaxation time. As the relaxation time is further increased,  the cells become less localized to the underlying rollers, and their now irregular oscillations reflect the presence of many frequencies. 

\end{abstract}	
% main text
\section{Introduction}
We study the dynamics of a layer of viscoelastic fluid that is driven from beneath by a 4x4 array of counter-rotating disks. The disks are arranged so that each adjacent pair counter-rotates at a given frequency $f$.  The videos 
(\href{http://ecommons.library.cornell.edu/bitstream/1813/14109/3/mpeg-1.mpg}{low-res.} and
\href{http://ecommons.library.cornell.edu/bitstream/1813/14109/2/mpeg-2.mpg}{high-res.}) show observations of fluids of different relaxation times driven by the disk array at the fixed spinning frequency $f=7.5$ Hz. The viscoelastic fluids are polyacrylamide (PAA)/aqueous-glycerol mixtures at different concentrations of PAA. The relaxation times $\lambda$ of the fluids are $0.04$s, $0.2$ s, and $2.6$s, respectively, obtained from measuring the relaxation time of the deformed free surface once the disks stop spinning.   

The relative forcing can be characterized by a Weissenberg number Wi, determined by the product of a shear rate $\dot\gamma$ to the relaxation time of the fluid $\lambda$. Here, the Reynolds number is below $1$, suggesting that fluid elasticity dominates, rather than inertia. By increasing the Weissenberg number, the fluid can be brought from a state of steady flow to one with oscillations, and then to one with very complex spatial and temporal dynamics. Our observations are related to other studies of low Reynolds number ``elastic turbulence" (e.g., Groisman \& Steinberg, {\it Nature} 2001; Thomases \& Shelley, {\it PRL} 2009), and suggest that the nonlinearities of polymer stress transport can play a role similar to those in the Newtonian Navier-Stokes equations in producing flows of spatial and temporal complexity.

\end{document}